\input harvmac

\Title{\vbox{\baselineskip12pt\hbox{UICHEP-TH/96-16}}}
{\vbox{\centerline{Relating Quarkonium Wave Functions at the Origin}}}

\centerline{Sterrett J. Collins,\foot{lobo@uic.edu}
Tom D. Imbo,\foot{imbo@uic.edu} B. Alex King\foot{aking@uic.edu} and
Eric C. Martell\foot{ecm@uic.edu}}

\bigskip\centerline{Department of Physics}
\centerline{University of Illinois at Chicago}
\centerline{845 W. Taylor St.}
\centerline{Chicago, IL \ 60607-7059}

\vskip 1.0in

Within the context of nonrelativistic potential models, we obtain several
formulas (with varying degrees of rigor) relating the wave functions at the
origin of the $c{\bar c}$, $b{\bar c}$ and $b{\bar b}$ S-wave quarkonium
systems. One of our main results is a model-independent relation which seems
to hold to within 3\% for any reasonable choice of interquark potential and
any choice of radial quantum number --- namely, ${|\Psi_{b{\bar c}}(0)|^2
\simeq |\Psi_{c{\bar c}}(0)|^{1.3}|\Psi_{b{\bar b}}(0)|^{0.7}}$ (the exponents
are motivated in the text). One of the physical consequences of this result is
the following relationship between heavy meson masses which we expect to hold
at about the 10\% level: ${M_{B_c^*}-M_{B_c}\simeq (0.7)(M_{J/\psi}
-M_{\eta_c})^{0.65}(M_{\Upsilon}-M_{\eta_b})^{0.35}}$.

\Date{}

Recent major advances in our understanding of the nonrelativistic limit of
Quantum Chromodynamics (QCD) have generated renewed interest in the
calculation of the production and decay rates of heavy quark bound states
\ref\nr{G.~T.~Bodwin, E.~Braaten and G.~P.~Lepage, Phys. Rev. D {\bf 51}
(1995) 1125, and references therein.}. However, these computations still
contain numerous nonperturbative ``parameters'' which cannot as yet be
accurately determined analytically from first principles. They must either be
fit to experiment, determined numerically from lattice simulations of QCD, or
extracted by more phenomenolgical considerations (for example, from potential
models). One important class of such parameters is the wave function at the
origin (WFO), $\Psi (0)$, for an S-wave bound state of a heavy quark and
anti-quark. (More generally, for a bound state with angular momentum $\ell$,
we should consider the quantity $d^{\ell}\Psi /dr^{\ell}$ evaluated at the
origin. However, in what follows we will concentrate on the case $\ell=0$.)
The WFO enters not only into the production and decay amplitudes for heavy
quarkonium systems, but also into the determination of the hyperfine splitting
in their mass spectra. Large QCD and relativistic corrections to the simple
first order formulas which relate the WFO's to the above observables
(especially for charmonium) make it difficult to extract precise information
about the WFO's from the experimental data. Moreover, while there has been
much recent progress (see, for example, \ref\lat{G.~T.~Bodwin, D.~K.~Sinclair
and S.~Kim, Phys. Rev. Lett. {\bf 77} (1996) 2376.}), lattice simulations of
QCD are not yet accurate enough to be very useful. This leaves us at the mercy
of a more model-dependent approach.

In nonrelativistic potential model descriptions of heavy quarkonia, it is a
simple numerical exercise to extract highly accurate values of the WFO's for
any given choice of the static potential between a heavy quark and anti-quark.
The problem lies in which potential to choose. Two potentials which yield very
similar spectra for the heavy mesons can give very different WFO's for any
given state. The WFO's seem to be quite sensitive to the global details of the
potential, while the energy levels are only sensitive to the shape of the
potential in the vicinity of the RMS radii of the states being studied. In
other words, when this sensitivity of the WFO's is coupled with our ignorance
regarding the details of the heavy quark potential, we again arrive at the
sobering conclusion that an accurate determination of the WFO's seems beyond
our reach.

So what's a theorist to do? Well, the situation isn't quite so dire as we have
made it sound. There are certain rigorous, {\it qualitative} statements that
one can make about WFO's within the context of potential models which hold for
very large classes of potentials --- including those believed to be relevant
for heavy quark systems. For example, in a {\it concave downward} potential
$V(r)$ (defined by $V^{\prime}(r) > 0$ and $V^{\prime\prime}(r) < 0$ for all
values of the interquark distance $r$), it can be shown that the square of the
WFO for the 1S state is larger than that of the 2S state \ref\art{A.~Martin,
Phys. Lett. {\bf 70B} (1977) 192.}. (Most likely it is more generally true
that the square of the WFO decreases monotonically with increasing radial
quantum number, but this has not, to our knowledge, been shown yet.) Moreover,
it can be proven from lattice QCD that the static potential between two heavy
color sources {\it is} concave downward \ref\con{C.~Borgs and E.~Seiler, Comm.
Math. Phys. {\bf 91} (1983) 329\semi C.~Bachas, Phys. Rev. D {\bf 33} (1986)
2723.}. An accurate enough determination of the WFO's for low-lying heavy
mesons can be made from the experimental data in order to test this
qualitative result, despite the size of QCD and relativistic corrections. And
indeed it holds true in both the $c{\bar c}$ and $b{\bar b}$ systems. Another
such result is that the square of the WFO for the 1S state increases with the
two-body reduced mass $\mu$ faster than linearly for a concave downward
potential \ref\qr{C.~Quigg and J.~L.~Rosner, Phys. Rep. {\bf 56} (1979)
167\semi H.~Grosse and A.~Martin, Phys. Rep. {\bf 60} (1980) 341.}. (This
result is also conjectured, but not proven, for higher states.) A comparison
of the WFO's extracted from the $c{\bar c}$ and $b{\bar b}$ data is consistent
with this theorem as well.

Should we have panicked if either of these two results were violated by the
WFO's obtained from the data? No, not necessarily. We could have attributed
the discrepancy to (at least) two possible sources. First, the uncertainty in
any WFO extracted from the data can be estimated from an educated guess at the
size of the higher order QCD and relativistic corrections not included in the
determination (as well as the experimental uncertainty in the measurement). It
is possible that these corrections are larger than we expect, and therefore
a WFO obtained with the truncated series is not as accurate as we thought.
Second, it is possible that our naive picture of, say, the $J/\psi$ as a
simple bound state of a $c$ quark and a ${\bar c}$ quark in a relative
S-wave interacting via a static potential is incorrect. For example, one may
have to consider a dynamical treatment of excited glue inside the meson, or
allow mixing with other angular momentum states and/or continuum states. There
is a nice example of this latter possibility \ref\wr{D.~B.~Lichtenberg, Phys.
Rev. D {\bf 49} (1994) 6244.}. It is strongly believed that in a concave
downward potential, the energy splitting between the $(n+2)$S and $(n+1)$S
states is always less than the splitting between the $(n+1)$S and $n$S states,
for any $n\geq 1$. However, in the charmonium system, though the measured
3S-2S splitting is less than the 2S-1S splitting, the 4S-3S splitting shows
an increase over the 3S-2S difference. Accepting the truth of the above
conjecture concerning energy splittings in a concave downward potential, how
do we explain the experimental numbers? The answer is that the threshold for
open charm production occurs between the 2S and 3S levels and induces
substantial mixing of the 3S and 4S $c{\bar c}$ states with continuum states.
Thus, what we experimentally identify as the 3S and 4S levels of charmonium
actually have substantial $D{\bar D}$ and $D^*{\bar D^*}$ components, among
others. A similar state of affairs occurs in the bottomonium system. Here,
the measured 5S-4S splitting is greater than the 4S-3S splitting. The open
bottom threshold occurs between the 3S and 4S levels and causes substantial
mixing of the 4S and 5S $b{\bar b}$ states with continuum states. It is this
mixing in both the $c{\bar c}$ and $b{\bar b}$ systems that seems to be
responsible for the apparent violations of the above energy splitting
 conjecture.
This interpretation is supported by a coupled-channel analysis in the charmonium
system \ref\coup{M.~Hirano, T.~Honda, K.~Kato, Y.~Matsuda and M.~Sakai, Phys.
Rev. D {\bf 51} (1995) 2353.}. The only purpose in showing this example is to
remind the reader that no matter how generally a certain result may apply
within the context of nonrelativistic potential models, there are still
assumptions that must be made in order to relate these potential model results
to real observations. And these assumptions may not hold for all states in all
systems.

With this disclaimer behind us, we can now describe the results of this note.
The starting point for our investigation is a recent paper by Eichten and
Quigg \ref\eq{E.~J.~Eichten and C.~Quigg, Phys. Rev. D {\bf 52} (1995) 1726.}
which tabulates the WFO's for various quarkonium states in an assortment
of ``successful'' potential models. We list them below (in natural units).
\vskip 12pt
\noindent
{\bf (1)} The Martin potential \ref\mar{A.~Martin, Phys. Lett. {\bf 93B}
(1980) 338.}: $V(r)=Ar^{0.1}+C$, where $A=6.898\ {\rm GeV}^{1.1}$,
$m_c=1.8\ {\rm GeV}$ and $m_b=5.174\ {\rm GeV}$.
\vskip 12pt
\noindent
{\bf (2)} The log potential \ref\log{C.~Quigg and J.~L.~Rosner, Phys. Lett.
{\bf 71B} (1977) 153.}: $V(r)=A\ell n(r/r_0)$, where $A=0.733\ {\rm GeV}$,
$m_c=1.5\ {\rm GeV}$ and $m_b=4.906\ {\rm GeV}$.
\vskip 12pt
\noindent
{\bf (3)} The Cornell potential \ref\cor{E.~Eichten, K.~Gottfried,
T.~Kinoshita, K.~D.~Lane and T.-M.~Yan, Phys. Rev. D {\bf 21} (1980) 203.}:
$V(r)=-A/r+Br+C$, where $A=0.52$, $B=(1/2.34)^2\ {\rm GeV}^2$, $m_c=1.84\
{\rm GeV}$ and $m_b=5.17\ {\rm GeV}$.
\vskip 12pt
\noindent
{\bf (4)} The Buchm\"uller-Tye potential \ref\bt{W.~Buchm\"uller and
S.-H.~H.~Tye, Phys. Rev. D {\bf 24} (1981) 132.}: This potential has a rather
complicated position space form. It is linear at large distances and
quasi-Coulombic at short distances. The deviations from pure Coulombic
behavior reproduce the running of the strong coupling constant to
next-to-leading order in QCD. The global shape of the potential is essentially
determined by two parameters --- namely, the QCD scale (in the modified
minimal subtraction scheme) $\Lambda_{\overline{MS}}$ which the authors of
\bt\ fit to be $509\ {\rm MeV}$, and the QCD string tension which they take
to be $0.153\ {\rm GeV}^2$ (motivated by the light meson data). The potential
also depends on the number of ``light'' flavors $n_f$. The authors take
$n_f=3$ for $r\geq 0.01\ {\rm fm}$, and $n_f=4$ for $r < 0.01\ {\rm fm}$. The
quark masses used are $m_c=1.48\ {\rm GeV}$ and $m_b=4.88\ {\rm GeV}$.
\vskip 12pt
\noindent
(The parameters $C$ in (1) and (3) and $r_0$ in (2) are irrelevant for $|\Psi
(0)|^2$.) Eichten and Quigg treat the $c{\bar c}$, $b{\bar b}$ and $b{\bar c}$
systems.

The first thing that catches one's eye in glancing at these tables is the
apparent randomness of the entries. Of course one can spot the aforementioned
general trends --- namely, for a fixed quark content, the square of the S-wave
WFO decreases with increasing radial excitation, and for fixed quantum numbers
the square of the WFO gets bigger as one goes from the $c{\bar c}$ to the
$b{\bar c}$ to the $b{\bar b}$ system (increasing reduced mass). However,
besides these qualitative behaviors, no additional regularity is apparent. For
example, the square of the WFO for the $\Upsilon$ changes by about a factor of
3 between the various potentials --- potentials which yield basically the same
low-lying spectrum! Things like the ratio of the $\psi (2S)$ and the $J/\psi$
WFO's, or the ratio of the $\Upsilon$ and the $J/\psi$ WFO's, also cover a
large range of values. Can any additional statements about these numbers be
made which possess some degree of model-independence?

Before we address this question, we would first like to present our own
version of the S-wave portion of Tables I-III in \eq , which corrects some
small numerical errors made there. For instance, it is well known that for
power-law potentials $V(r)=Ar^a+C$, the square of the S-wave WFO scales with
reduced mass $\mu$ as $\mu^{3/(2+a)}$ \qr . This result can also be used for
the log potential by putting $a=0$. The results of \eq\ show a mild violation
of this scaling (on the order of a few percent for all radial quantum numbers)
which cannot be accounted for by rounding errors. Upon our redoing of the
computations using the Runge-Kutta method for solving the nonrelativistic
Schr\"odinger equation, we found results which satisfied the scaling laws
(within rounding errors) for the log and Martin potentials, and typically
disgreed with the results of \eq\ in the second significant figure. We also
tested our program on potentials with analytically known WFO's, such as the
Coulomb, linear and harmonic oscillator potentials, and obtained agreement
with the exact results to at least six significant figures. We
then ran our program on the other potentials treated in \eq , the Cornell and
B\"uchmuller-Tye potentials, and found similar disagreements to those
encountered in the log and Martin cases. It should be stressed however that
these mild errors in no way affect the conclusions of \eq . We just want
numerical results which are as accurate as possible in order to test some
approximate formulas relating different WFO's that we will derive later.

For all of these potentials we display results for the ground state as well as
the next five radial excitations. This goes a little further than the results
in \eq . Many of these states lie above the threshold for open flavor
production, and hence in a region where the WFO's have limited usefulness
because of mixing with continuum states. However, these numbers are still
quite useful in checking the general validity of the analytic formulas which
are to come.

We have also added one additional potential to the table.
\vskip 12pt
\noindent
{\bf (5)} The Lichtenberg-Wills potential \ref\lw{D.~B.~Lichtenberg and
J.~G.~Wills, Nuovo Cimento {\bf 47A} (1978) 483.}: $V(r)=8\pi (1-\Lambda r)^2/
[(33-2n_f) r\ell n(\Lambda r)]$, where we choose $\Lambda =0.7\ {\rm GeV}$,
$m_c=1.84\ {\rm GeV}$ and $m_b=5.17\ {\rm GeV}$. At short distances, the
running of the strong coupling constant to leading order in QCD is reproduced
if one identifies $\Lambda =e^{\gamma}\Lambda_{QCD}$, where $\gamma$ is
Euler's constant. We have also taken the number of light flavors $n_f$ to be
three in all systems studied with this potential.
\vskip 12pt
\noindent
Note that we have chosen the $b$ and $c$ constituent quark masses to be the
same as for the Cornell potential. The parameter $\Lambda$ was then chosen so
as to obtain a low-lying meson spectrum reasonably close to that obtained from
the Cornell potential. The differences in the WFO's between the Cornell and
Lichtenberg-Wills examples are then basically due to the different shapes
of the potentials outside of the region between the RMS radii of the
$c{\bar c}$ and $b{\bar b}$ systems. A substantial difference can still be
seen between the two sets of WFO's, again emphasising their sensitivity to
global features of the interquark potential.

We first became interested in finding regularities in these numbers after a
comment made to one of us by Ira Rothstein. He was able to prove that, in
nonrelativistic QCD in the limit as $m_b\to m_c$, the square of the WFO for
any state in the $b{\bar c}$ system is equal to the average of the squares of
the WFO's of the corresponding $c{\bar c}$ and $b{\bar b}$ states, plus a
correction of order $\delta^2$ where $\delta=(m_b-m_c)/(m_b+m_c)$
\ref\ira{I.~Z.~Rothstein, private communication.}. That is,
\eqn\one{|\Psi_{b{\bar c}}(0)|^2=(|\Psi_{c{\bar c}}(0)|^2+|\Psi_{b{\bar b}}
(0)|^2)/2 + O(\delta^2).}
We have shown that this is also true in an arbitrary nonrelativistic
potential model. Indeed, one can prove a slightly stronger result:
\eqn\two{|\Psi_{b{\bar c}}(0)|=(|\Psi_{c{\bar c}}(0)|+|\Psi_{b{\bar b}}
(0)|)/2 + O(\delta^2).}
A similar formula also holds for the geometric mean instead of the arithmetic
mean:
\eqn\three{|\Psi_{b{\bar c}}(0)|^2=|\Psi_{c{\bar c}}(0)||\Psi_{b{\bar b}}
(0)|+ O(\delta^2).}
These last two results can be easily demonstrated from perturbation theory
in $\delta$. However, even though these results are independent of the nature
of the interquark forces, they are unfortunately not very useful in real
applications since the quantity $\delta$ is approximately 1/2 for reasonable
values of $m_b$ and $m_c$. The order $\delta^2$ corrections in the above
equations are therefore large, which a simple check using the numbers in Table
I will show.

What we want is a relation with the model-independence of Eqs.(1)-(3), but
with much more quantitative accuracy. For the class of power-law potentials
$V(r)=Ar^a+C$ discussed earlier, there is a very simple, exact relationship
between the WFO's of the $c{\bar c}$, $b{\bar c}$ and $b{\bar b}$ systems.
In order to derive this relation, we first recall that simple scaling
arguments for the above power-law potentials tell us that for any reduced mass
$\mu$ we have ${|\Psi_{\mu}(0)|^2=f(n,a)(A\mu )^{3/(2+a)}}$, where $f(n,a)$ is
only a function of the radial quantum number $n$ and the power $a$. Using this
fact alone, it is straightforward to obtain, for reduced masses $\mu_1 < \mu_2
< \mu_3$ and any fixed $n$,
\eqn\four{|\Psi_{\mu_2}(0)|^2=|\Psi_{\mu_1}(0)|^{2(1-q)}|\Psi_{\mu_3}(0)|
^{2q},}
where $q=\ell n(\mu_2 /\mu_1)/\ell n(\mu_3 /\mu_1)$. Choosing $\mu_1 =m_c/2$,
$\mu_2=m_bm_c/(m_b+m_c)$ and $\mu_3=m_b/2$, this becomes
\eqn\five{|\Psi_{b{\bar c}}(0)|^2=|\Psi_{c{\bar c}}(0)|^{2(1-q)}|\Psi_
{b{\bar b}}(0)|^{2q},}
where $q=\ell n(2m_b/(m_b+m_c))/\ell n(m_b/m_c)$. This result is nice not only
for its simplicity, but also because it does not depend on any of the
parameters ($A$, $a$ and $C$) appearing in the potential. It depends only on
the constituent quark masses $m_b$ and $m_c$. It is easy to check this result
on the log ($a=0$) and Martin ($a=0.1$) potentials in Table I. Since this
formula has no dependence on parameters in the potential, we can also check it
on the other examples in Table I. Of course it will no longer be exact in
these cases since the above scaling law for $|\Psi_{\mu}(0)|^2$ is true (for
all $\mu$) only for power-law potentials. And these other potentials are far
from being power-like. They each have a (quasi-)Coulombic nature at small $r$,
motivated from one gluon exchange, and a (quasi-)linear behavior at large $r$,
motivated by a stringy picture of confinement. In this sense, they are more
``realistic'' than the power-law potentials. In the intermediate $r$ range
containing the RMS radii of the heavy quarkonium states, they are
quasi-logarithmic, just like the log and Martin potentials. But, though
Eq.(5) is not exact here, we can still ask if it is a reasonably accurate
approximation.

The answer is yes. For every choice of $n$ in Table I, the relation in Eq.(5)
holds to within 4\%\ (except for the 1S state of the Cornell potential where
it is off by about 7\% ). The least accurate results are obtained for the
ground state. As $n$ increases, the results get better. This is a substantial
improvement over the accuracy of Eqs.(1)-(3). The only price that we have had
to pay is the introduction of the constituent quark masses into the relation.
It is interesting to note that the left-hand side of Eq.(5) is less than or
equal to the right-hand side for each potential considered and each choice of
$n$. Is it possible that this is always the case --- at least for a wide class
of potentials? A numerical study of numerous examples, as well as an analysis
of the question within the context of various approximation schemes, has led
us to the following conjecture:
\vskip 12pt
\noindent
{\bf Conjecture:} Consider a potential $V(r)$ such that $p(r)\equiv
rV^{\prime\prime}(r)/V^{\prime}(r)$ is monotonically increasing with
increasing $r$. Then for each choice of radial quantum number, and for reduced
masses $\mu_1 < \mu_2 < \mu_3$, we have
\eqn\six{|\Psi_{\mu_2}(0)|^2<|\Psi_{\mu_1}(0)|^{2(1-q)}|\Psi_{\mu_3}(0)|
^{2q},}
where $q$ is as in Eq.(4). For $p(r)$ monotonically decreasing with increasing
$r$, the inequality in Eq.(6) is reversed.
\vskip 12pt
\noindent
Of course when $p(r)$ is independent of $r$, $V(r)$ is a power-law potential
and the inequality in Eq.(6) is replaced by the equality of Eq.(4). One can
think of $1+p(r)$ as the ``effective power'' of $V(r)$ at quark separation
$r$. We will call a potential {\it power increasing}, or PI, when $p(r)$ is
monotonically increasing, and {\it power decreasing}, or PD, when $p(r)$ is
monotonically decreasing. Each of the non-power-law quarkonium potentials in
Table I is PI (we have checked this numerically for the Buchm\"uller-Tye
potential), and satisfies the inequality in Eq.(6) with the appropriate
choices of $\mu_1$, $\mu_2$ and $\mu_3$ --- namely
\eqn\seven{|\Psi_{b{\bar c}}(0)|^2<|\Psi_{c{\bar c}}(0)|^{2(1-q)}|\Psi_
{b{\bar b}}(0)|^{2q},}
where $q$ is as in Eq.(5). Indeed, all of the popular potentials used in
quarkonium studies seem to be PI. But, unlike the concave downward property,
we know of no QCD-motivated reason why this must be so. But we conjecture
that Eq.(7) holds in the nonrelativistic limit of QCD, and in all realistic
potential models.

Actually, there is an even better result which is {\it completely}
parameter-independent. To obtain this, we first note that $q=\ell n(2m_b/
(m_b+m_c))/\ell n(m_b/m_c)$ lies between about 0.36 and 0.38 for any
reasonable choices of $m_b$ and $m_c$. However, as noted above, when
substituted into Eq.(5) this yields WFO's for the $b{\bar c}$ system which are
too high. In the context of the general form of Eq.(5), the potentials of
interest seem to favor a slightly lower value of $q$. We have found that if
$q$ is simply taken to be 0.35 independent of the interquark potential and
quark masses being considered, very accurate results are obtained. That is,
we have
\eqn\eight{|\Psi_{b{\bar c}}(0)|^2\simeq |\Psi_{c{\bar c}}(0)|^{1.3}|\Psi_
{b{\bar b}}(0)|^{0.7}.}
Although no longer exact for power-law potentials, this simple formula holds
to within 2.5\% for all cases in Table I. This is quite remarkable given the
range of radial quantum numbers covered and the global differences in the
potentials treated. We fully expect it to have a similar accuracy for any
reasonable quarkonium potential. Though not on the same rigorous footing as
the two qualitative theorems discussed earlier, it is reasonably well
motivated by Eqs.(5) and (7) above. Moreover, it gives us a better
{\it quantitative} understanding of the jumble of numbers in Table I.

Can we extract any simple physical consequences of this result? Certainly it
implies relationships between the production (and decay) amplitudes for the
$J/\psi$, $B_c$ and $\Upsilon$ systems. However, it is perhaps simpler to
discuss the implications for the hyperfine mass splittings in these systems.
To leading order in $\alpha_s$ and $v^2/c^2$, the mass splitting $(\Delta M)
_{{\rm i{\bar j}}}$ (for any fixed $n$) between the vector and pseudoscalar
mesons composed of a quark of flavor i and an antiquark of flavor j (of mass
$m_{{\rm i}}$ and $m_{{\rm j}}$, respectively) is given by
\eqn\nine{(\Delta M)_{{\rm i{\bar j}}}=32\pi\alpha_s(2\mu_{{\rm i{\bar j}}})
|\Psi_{{\rm i{\bar j}}}(0)|^2/9m_{{\rm i}}m_{{\rm j}},}
where $\mu_{{\rm i{\bar j}}}=m_{{\rm i}}m_{{\rm j}}/(m_{{\rm i}}+
m_{{\rm j}})$, and we have assumed the standard Breit-Fermi hyperfine
interaction \ref\hf{W.~Lucha, F.~F.~Sh\"oberl and D.~Gromes, Phys. Rep.
{\bf 200} (1991) 270.}. Putting this together with Eq.(8) gives
\eqn\ten{(\Delta M)_{b{\bar c}}=\alpha_s(2\mu_{b{\bar c}})(m_c/m_b)^{0.3}
[(\Delta M)_{c{\bar c}}/\alpha_s(m_c)]^{0.65}[(\Delta M)_{b{\bar b}}/
\alpha_s(m_b)]^{0.35}.}
It is interesting to note that for any reasonable choices of $m_b$, $m_c$,
and $\Lambda_{QCD}$, the quantity ${\alpha_s(2\mu_{b{\bar c}})/\alpha_s(m_c)
^{0.65}\alpha_s(m_b)^{0.35}}$ has a numerical value which is within about 3\%
of 1. (In a similar fashion, both QCD and relativistic corrections to Eq.(9)
approximately cancel when fed into Eq.(10).) Moreover, $(m_b/m_c)^{0.3}$ is
always within a few percent of 0.7. Therefore, we can write
\eqn\eleven{(\Delta M)_{b{\bar c}}\simeq (0.7)(\Delta M)_{c{\bar c}}
^{0.65}(\Delta M)_{b{\bar b}}^{0.35}.}
For the ground state, this reads
\eqn\twelve{M_{B_c^*}-M_{B_c}\simeq (0.7)(M_{J/\psi}-M_{\eta_c})
^{0.65}(M_{\Upsilon}-M_{\eta_b})^{0.35}.}
Given the sources of uncertainty enumerated above, we expect this result to
hold at about the 10\% level. Only three of the six mesons appearing in
Eq.(12) have been found experimentally thus far --- namely, the $J/\psi$,
$\eta_c$ and $\Upsilon$. Their masses are known quite accurately
\ref\pdg{Particle Data Group, Phys. Rev. D {\bf 54} (1996) 1.}. There is some
hope that the remaining three mesons will be detected in the near future at
either the Fermilab Tevatron or LEP, allowing a test of the above result.

\vskip 12pt
{\centerline {\bf Acknowledgements}}
\vskip 12pt
\noindent
A special thank you to Ira Rothstein for starting us thinking about these
issues. It is also a pleasure to thank David Bergmann, David Bowser-Chao and
Adam Falk for useful discussions. This work was supported in part by the
U.~S.~Department of Energy under contract number DE-FG02-91ER40676.

\listrefs

\offinterlineskip
\indent\halign{\vrule\quad#\hfil\vrule&\strut\quad\hfil#\hfil\quad
&\vrule\quad#\hfil\quad\vrule
&\quad#\hfil\quad\vrule
&\quad#\hfil\quad\vrule
&\quad#\hfil\quad\vrule
&\quad#\hfil\quad\vrule
&\quad#\hfil\quad\vrule
\cr
\noalign{\hrule}
Potential               &System &\multispan6\vrule\hfill $|R(0)|^2$
 \hfill\vrule\cr
&\omit &\multispan6\unskip\leaders\hrule\hfill\vrule\cr
& &1S     &2S     &3S     &4S     &5S     &6S     \cr
\noalign{\hrule}
Martin                &$c\bar{c}$     &0.979  &0.545  &0.390  &0.309  &0.257
 &0.222  \cr
			&$b\bar{c}$     &1.720  &0.957  &0.685  &0.542  &0.452  &0.390  \cr
			&$b\bar{b}$     &4.423  &2.461  &1.763  &1.394  &1.164  &1.004  \cr
\noalign{\hrule}
logarithmic             &$c\bar{c}$     &0.796  &0.406  &0.277  &0.211  &0.172
 &0.145  \cr
			&$b\bar{c}$     &1.508  &0.770  &0.524  &0.401  &0.325  &0.275  \cr
			&$b\bar{b}$     &4.706  &2.401  &1.636  &1.250  &1.015  &0.857  \cr
\noalign{\hrule}
Cornell                 &$c\bar{c}$     &1.458  &0.930  &0.793  &0.725  &0.683
 &0.654  \cr
			&$b\bar{c}$     &3.191  &1.769  &1.449  &1.297  &1.205  &1.141  \cr
			&$b\bar{b}$     &14.06  &5.681  &4.275  &3.672  &3.322  &3.088  \cr
\noalign{\hrule}
Buchm\"uller-Tye        &$c\bar{c}$     &0.794  &0.517  &0.441  &0.404  &0.381
 &0.365  \cr
			&$b\bar{c}$     &1.603  &0.953  &0.785  &0.705  &0.658  &0.625  \cr
			&$b\bar{b}$     &6.253  &3.086  &2.356  &2.032  &1.845  &1.721  \cr
\noalign{\hrule}
Lichtenberg-Wills       &$c\bar{c}$     &1.121  &0.693  &0.563  &0.496  &0.453
 &0.423  \cr
			&$b\bar{c}$     &2.128  &1.231  &0.975  &0.846  &0.766  &0.711  \cr
			&$b\bar{b}$     &6.662  &3.370  &2.535  &2.139  &1.902  &1.740  \cr
\noalign{\hrule}
       }

\vskip 24pt
\noindent
{\bf Table 1:} Numerical values of the radial wave function at the origin
squared, $|R(0)|^2=|\Psi (0)|^2/4\pi$, for the first six S-wave states of
heavy quarkonium systems in various potential models. The parameters used in
the potentials are discussed in the text.

\bye